# Trajectory-based cycle-level pedestrian crash prediction at intersections: A novel non-linear link function for block maxima led Bayesian GEV framework addressing heterogeneous traffic conditions

Parvez Anowar[a], Nazmul Haque[b], Md Asif Raihan[b,*] and Md Hadiuzzaman[c]

[a] *Department of Civil, Environmental and Construction Engineering, University of Central Florida, 12800 Pegasus Drive, Orlando, FL 32816, USA*

[b] *Accident Research Institute, Bangladesh University of Engineering and Technology, Dhaka 1000, Bangladesh*

[c] *Department of Civil Engineering, Bangladesh University of Engineering and Technology, Dhaka 1000, Bangladesh*

* Corresponding author: Md Asif Raihan (raihan@ari.buet.ac.bd)

E-mail addresses: pa545735@ucf.edu (P. Anowar), nhaque@ari.buet.ac.bd (N. Haque), raihan@ari.buet.ac.bd, mraih001@fiu.edu (M.A. Raihan), mhadiuzzaman@ce.buet.ac.bd (M. Hadiuzzaman)

## Abstract

This study develops a cycle-level framework for estimating pedestrian crash risk at signalized intersections under heterogeneous, non-lane-based traffic. Existing approaches often assume linear relationships between covariates and parameters, oversimplifying the complex, non-monotonic interactions among motorized vehicles (MVs), non-motorized vehicles (NMVs), and pedestrians. To address this limitation, the framework introduces a non-linear link function within a Bayesian generalized extreme value (GEV) structure to capture how traffic effects vary. It applies extreme value theory through the block maxima approach using post-encroachment time as a surrogate safety measure, with each signal cycle serving as an analysis block. It also proposes a behaviour-normalized modified crash risk (MRC) metric to account for pedestrians' habitual risk-taking behaviour in Dhaka's mixed traffic. Seven hierarchical Bayesian models were estimated using Markov Chain Monte Carlo simulation and compared using the deviance information criterion. Models with a non-linear link function only in the GEV location parameter fit the data better than their linear counterparts. Results suggest pedestrian speed is negatively associated with crash risk, while flow and speed of MVs, pedestrian flow, and NMV conflicting speed show positive associations, though all remain uncertain. The selected model's MRC estimate was close to the observed crashes at a 93% confidence level tuned to this dataset. These results suggest that the non-linear structure adds flexibility for capturing the non-stationary and non-linear nature of traffic extremes. This framework is a step towards proactive identification of pedestrian crash risk and after wider validation, it could support adaptive signal control and safety interventions in mixed-traffic environments.

## 1. Introduction

Proactive road safety assessment methodologies incorporating surrogate safety measures have gained momentum to overcome the limitations of historical crash data. Extreme Value Theory (EVT) [1], probabilistic frameworks [2], and causal models [3] have been widely used to estimate crash risks from traffic conflicts. EVT has seen increasing application because it can estimate crash risk from observable conflict events that precede rare or unobservable crashes. It offers an alternative to methods relying solely on police-reported data and avoids assumptions about fixed crash-to-surrogate ratios [4–6]. As a result, EVT has been extensively applied in developing crash risk models, particularly for vehicle-to-vehicle crashes.

EVT has also gained attention for real-time pedestrian crash risk assessment, where studies analyse sampling strategies, conflict indicators, traffic stream characteristics, link functions, and covariates to model crash risk (Table 1). These components guide how data and traffic events are selected and how relationships between variables and crash risk are represented. Covariates, in particular, help identify the main factors shaping pedestrian crash risk and support more effective safety interventions.

However, most studies have been limited to homogeneous traffic settings, with little consideration for mixed conditions where motorized vehicles (MVs) and non-motorized vehicles (NMVs) share the same carriageway. This gap can be attributed to study locations lacking significant NMV presence or having designated lanes. In heterogeneous conditions, differences in vehicle speeds, manoeuvrability, and pedestrian interactions produce varying levels of crash severity, influencing overall crash risk [7].

Table 1: Summary of some studies that use extreme value models

| Author | Sampling Approach | Conflict Indicator | Traffic Stream Type | MVs and NMVs Separated? | Link Function | Covariates |
|---|---|---|---|---|---|---|
| Tahir and Haque [8] | BM | PET and Delta-V | Hom. | No | Linear for LP and log-linear for SP | Number of vehicles, pedestrians, turning vehicles, and pedestrian conflicts per cycle; average and standard deviation of vehicle and pedestrian speeds per cycle |
| Ankunda et al. [9] | BM and POT | PET and Gap time | Het. | No | Log-linear for SP | Vehicle-pedestrian interactions count with PET ≤1.5 s, speed of pedestrians, speed of vehicles, group pedestrian crossing, buses, two-wheelers |
| Ali et al. [10] | BM | PET | Hom. | No | Linear for LP and log-linear for SP | Proportion of turning vehicle, pedestrian count, turning vehicle count, average speed of vehicles, and average speed of pedestrians |
| Ali et al. [11] | BM | Gap time | Hom. | No | Linear for LP and log-linear for SP | Speed, spacing, remaining distance, lag gap, and driver demographic |
| Fu and Sayed [12] | BM | MTTC | Hom. | No | Linear for LP and log-linear for SP | Traffic volume, shock wave area, and platoon ratio |

| Author | Sampling Approach | Conflict Indicator | Traffic Stream Type | MVs and NMVs Separated? | Link Function | Covariates |
|---|---|---|---|---|---|---|
| Alozi and Hussein [13] | POT | PET and TTC | Hom. | No | Log-linear for SP | conflict speed for vehicles, conflict speed for pedestrian, and vehicle movement type |
| Zhang and Abdel-Aty [14] | POT | PET and TTC | Hom. | No | Linear for SP | Pedestrian signal phase, cycle length, green time, red time, yellow time, and vehicle count during green light |
| Ali et al. [15] | POT | Gap time | Hom. | No | Log-linear for SP | Speed, spacing, remaining distance, lag gap, and driver demographic |
| Guo et al. [16] | POT | PET | Hom. | No | Log-linear for SP and linear for ShP | Vehicle's violation and pedestrian's violation |
| Zheng and Sayed [17] | BM | MTTC | Hom. | No | Linear for LP and log-linear for SP | Traffic volume, shock wave area, and platoon ratio |

**Abbreviations:** POT = Peak over threshold; PET = Post encroachment time; TTC = Time to collision; MTTC = Modified time to collision; Het. = Heterogeneous; Hom. = Homogeneous; LP = Location parameter; SP = Scale parameter; ShP = Shape parameter.

Prior research commonly assumes linear link functions for Generalized Extreme Value (GEV) parameters such as location and scale, which oversimplifies the complexity of traffic interactions. The theoretical basis for adopting non-linear link functions in extreme value models is well established. In road safety, Lao et al. [18] developed generalized non-linear models (GNM) for rear-end crash risk using Poisson formulations and demonstrated that generalized linear models (GLM) are a special case of GNMs. Their use of logarithmic and polynomial link functions showed that non-linear, non-monotonic relationships better represent real-world crash dynamics. Similarly, Carrodano [19] used a Bayesian network approach to identify non-linear interactions among human, vehicle, and environmental factors, emphasizing their interconnected influence on crash likelihood.

Beyond traffic applications, non-linear link functions have been widely used in hydrology, environmental science, oceanography, and climate research. Non-stationary GEV models have shown that quadratic, exponential, and spline-based functional forms more accurately describe changes in extremes than linear specifications [20–23]. Li et al. [24] advanced this framework through Bayesian hierarchical modelling of surface-level ozone extremes, demonstrating highly non-linear dependence of upper-tail behaviour on multiple covariates. In oceanography, Mackay and Jonathan [25] showed that non-stationary extreme value models with non-linear covariate relationships substantially improve return value estimates for storm peak behaviour. The evgam framework [26] unified these developments by providing a flexible generalized additive modelling structure for GEV and GPD parameters, where smooth non-linear functions define covariate effects and linear forms emerge as special cases.

Despite extensive use of non-linear approaches across disciplines, their application remains largely unexplored for block maxima (BM) crash risk estimation in GEV models. This study addresses this gap by comparing linear and non-linear link function specifications for vehicle-pedestrian crash risk. Through seven Bayesian hierarchical models incorporating different link forms, the research examines whether non-linear approaches improve model fit.

Additionally, many studies have focused on locations with relatively low numbers of conflicts, where crashes are less likely to occur, leaving high-conflict locations unaddressed. In such environments, conventional crash risk models often fail to account for pedestrians' habitual risk-taking behaviour, which can lead to an overestimation of predicted crashes compared to actual occurrences.

To address this overestimation in Dhaka's mixed, non-lane-based traffic, this study proposes a behaviour-normalized Modified Crash Risk (MRC) metric as an exploratory calibration strategy. In congested conditions, pedestrians frequently accept minimal time gaps, cross through moving traffic, and use conflict zones instead of designated facilities [27–31]. These practices, influenced by inadequate infrastructure and inconvenient overpasses, have normalized risk-taking in daily travel [32,33]. Based on Risk Homeostasis Theory [34,35], individuals adjust their behaviour to maintain a personal comfort level of risk, which may explain why conventional crash risk estimates remain high even under typical conditions. The proposed MRC incorporates a behavioural calibration intended to provide more context-sensitive safety assessments in Bangladesh's urban traffic settings. Its performance beyond the present study sites remains untested.

The overall contributions of this study are as follows:

(1) It develops a cycle-level vehicle-pedestrian crash risk framework for signalized intersections that considers both MVs and NMVs within the same traffic stream, applying a hierarchical Bayesian model using BM and GEV distribution.
(2) It incorporates both linear and non-linear link functions for GEV parameters and forms seven hierarchical models, comparing them through deviance information criterion (DIC) values to evaluate the performance of non-linear structures.
(3) It proposes a modified crash risk metric that accounts for behavioural adaptation and is intended to reduce the overestimation commonly observed in conventional crash risk formulations under mixed traffic conditions.

## 2. Methodology

Figure 1 presents the cycle-level crash risk estimation framework for vehicle-pedestrian interactions at signalized intersections proposed in this study. The framework was applied to pre-recorded videos and benchmarked against historical crash records.

### *2.1. Road user detection and trajectory extraction*

This study employed the DEEGITS framework [36] for vehicle and pedestrian detection, classification, and tracking using video sensors. The system utilized convolutional neural networks to identify road users under congested and occluded conditions. Calibration relied on the enriched "DhakaPersons" dataset, which applied data fusion for simultaneous vehicle and pedestrian detection. Image pre-processing and augmentation were used to improve dataset diversity, and transfer learning with a YOLOv8 pre-trained model enhanced detection accuracy across vehicle types. Hyperparameters were selected through Grid Search using the Stochastic Gradient Descent optimizer, and experiments confirmed high detection accuracy. DeepSORT was integrated for multi-object tracking under heterogeneous traffic. Tracked points were aggregated into vehicle trajectories and calibrated for perspective distortion using Equation (1) adopted from Hadiuzzaman et al. [37].

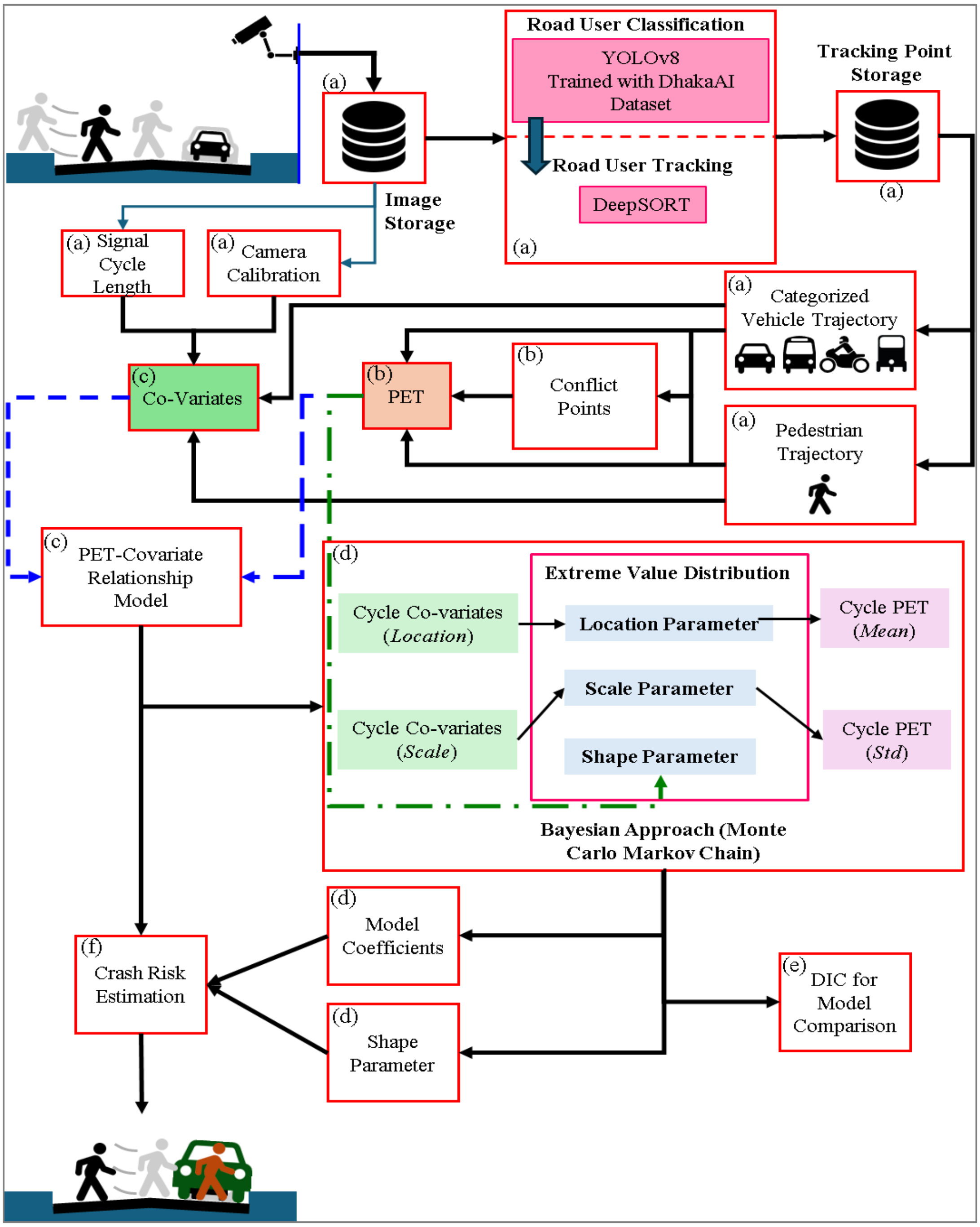

Figure 1. Signal cycle-level crash risk estimation framework for vehicle-pedestrian interactions

$$R = \theta\big(r, \rho(x, X)\big) \tag{1}$$

Where, $\rho(x, X)$ corrects the distorted distances $(r)$ due to decrease in length $X$ when a vehicle moves away at a distance x from the camera and $\theta$ converts the pictorial distance into field distance. The function $\rho$ can have any functional form. However, this study considers $\rho$ a 2nd-degree polynomial function to have the corrected distance $R$.

## 2.2. *Conflict identification and PET calculation*

Conflict points are identified by intersecting pedestrian and vehicle trajectories represented as polylines. When an intersection is detected, the intersecting coordinates $(x_{int}, y_{int})$ and timestamps of the pedestrian and vehicle tracks ($t_p$ and $t_v$) are recorded. PET is then computed based on these timestamps, as shown in Equation (2).

$$PET = f(x) = \begin{cases} t_v - t_p, & t_p < t_v \\ t_p - t_v, & t_v < t_p \\ 0 & t_v = t_p \end{cases} \tag{2}$$

PET values are subsequently aggregated over time intervals $\Delta t$. For each signal cycle $t_i$, the average PET and standard deviation are calculated to match the covariates described in Section 3.2.

## 2.3. *BM approach of EVT and incorporating covariates*

The proposed framework employs the BM approach of EVT to estimate crash risk from traffic conflicts. EVT is particularly well-suited for this purpose due to its ability to extrapolate from frequent traffic conflicts to rare crashes [1]. The signal cycle is used as the analysis interval because it captures dynamic interactions and risks associated with multiple road users at intersections [38] and fits naturally within the BM structure of the GEV distribution

In the BM approach, the highest value within each block of observations is considered an extreme value [39]. Mathematically, it involves a sequence of random and independent variables $(p_1, p_2, p_3, \dots\dots, p_n)$, each following a common distribution function, and $M_n = max(p_1, p_2, p_3, \dots\dots, p_n)$ denoting the block maximum of these values. Within this framework, $p_i$ represents a traffic conflict indicator such as PET measured during signal cycle $i$. Assuming that the maximum conflict values follow a GEV distribution as $n$ approaches infinity [40], the GEV distribution function can be expressed mathematically as Equation (3).

$$G(z) = \exp\left(-\left[1 + \xi\left(\frac{z-\mu}{\sigma}\right)\right]^{-\frac{1}{\xi}}\right) \tag{3}$$

Here, $z$ represents the observed or possible extreme value of the conflict indicator, $-\infty < \mu < \infty$ represents the location parameter, $\sigma > 0$ indicates the scale parameter and $-\infty < \xi < \infty$ represents the shape parameter. Equation (3) is used to describe traffic extremes (near-crash situations) while ensuring that the scale parameter remains strictly positive. It is parameterized as $GEV(\mu, \phi, \xi)$, where $\phi = \log \sigma$. Let, $p_{ij}$ be the $i^{th}$ cycle maximum for site $j$, where $j = 1, 2, 3, \dots\dots, s$ and $i = 1, 2, 3, \dots\dots, n_j$. That means $p_{ij}$ represents the maximum value for cycle $i$ and site $j$ and the GEV distribution, which is site-dependent, can be expressed as Equation (4):

$$G(z_{ij} < z | \mu_{ij}, \phi_{ij}, \xi_{ij}) = \exp\left(-\left[1 + \xi_{ij}\left(\frac{p - \mu_{ij}}{exp(\phi_{ij})}\right)\right]^{-\frac{1}{\xi_{ij}}}\right) \tag{4}$$

To represent temporal variation in extremes, covariates are incorporated through identity link functions. However, accurately estimating the shape parameter can be challenging due to the varied distribution of conflict extremes and the absence of covariates [39]. When GEV parameters are related to covariates, the process is non-stationary; otherwise, it is referred to as a stationary process ($\alpha_{\mu 1} = \alpha_{\phi 1} = 0$). The non-stationary process is represented by Equation (5) as follows:

$$\begin{pmatrix} \mu_{ij} = \alpha_{\mu 0} + \alpha_{\mu 1}\mathrm{X}^{\theta_\mu} + \varepsilon_{\mu j} \\ \phi_{ij} = \alpha_{\phi 0} + \alpha_{\phi 1}\mathrm{Y}^{\theta_\phi} + \varepsilon_{\phi j} \\ \xi_{ij} = \alpha_{\xi 0} + \varepsilon_{\xi j} \end{pmatrix} \tag{5}$$

In the equation above, $\alpha_{\mu 0}$, $\alpha_{\phi 0}$, and $\alpha_{\xi 0}$ represent intercept terms, and $\alpha_{\mu 1}$ and $\alpha_{\phi 1}$ denote parameter estimates for the covariate vectors X and Y. $\varepsilon_{\mu j}$, $\varepsilon_{\phi j}$, and $\varepsilon_{\xi j}$ stand for random error terms. $\theta_\mu$ and $\theta_\phi$ represent the exponent of the covariates of the location parameter and scale parameter respectively. It is worth noting that the random terms in Equation (5) signify variances between sites, which remain constant for extremes at the same site but vary across different sites. When $\theta_\mu = \theta_\phi = 1$, $\mu_{ij}$ and $\phi_{ij}$ will indicate linear relationship among the covariates. Previous studies [8–10,12] primarily adopted such linear forms, which tend to oversimplify the link functions and fail to reflect the non-monotonic nature of real-world traffic dynamics.

Introducing non-linear link functions through the exponent term $\theta$ provides greater flexibility in modelling complex, non-additive relationships involving traffic flow, pedestrian movement, and vehicle heterogeneity. Non-linear structures can capture thresholds, saturation effects, and varying covariate influence without requiring additional statistical tests [41]. This can make the non-linear GEV formulation more suitable for heterogeneous, mixed-traffic environments.

## 2.4. Bayesian approach for model estimation

To estimate the model in Equations (4) and (5), a Bayesian approach is used because it allows posterior estimation of parameters through specified priors. The model is structured by assigning priors to seven fundamental parameters ($\alpha_{\mu 0}$, $\alpha_{\phi 0}$, $\alpha_{\xi 0}$, $\alpha_{\mu 1}$, $\alpha_{\phi 1}$, $\theta_\mu$, and $\theta_\phi$). The first five parameters are assumed to have uninformative priors, modelled with a normal distribution having a mean of zero and a variance of $10^6$ (N (0, $10^6$)) suggested by Ntzoufras [42]. The last two parameters, $\theta_\mu$ and $\theta_\phi$, are considered to follow a uniform distribution within the range of (-2.0, 2.0) to avoid unwanted simulation environment (i.e. logarithm of negative value). However, estimating the GEV shape parameter can lead to convergence difficulties if priors are too restrictive [4]. To address this, the shape parameter is assigned a uniform prior over (-5.0 to 5.0) which is slightly larger than the previous studies [1,4]. Posterior distributions of all parameters are obtained using Markov Chain Monte Carlo (MCMC), with Gibbs sampling applied for model estimation.

## 2.5. Measure of performance

In Bayesian analysis, the DIC is often utilized for model comparison, offering a common goodness-of-fit measure [43] and it is expressed as in Equation (6).

$$DIC = \overline{D} + p_D \tag{6}$$

Here, $\overline{D}$ represents the posterior mean deviance, which evaluates how well the model fits the data, while $p_D$ indicates the effective number of parameters in the model. Generally, a lower DIC value suggests a better fit for the model. When considering multiple covariates, various models are created and compared, and the model with the lowest DIC value is selected as the best fit.

## 2.6. Crash risk estimation

The crash risk for each signal cycle was estimated using the fitted GEV distribution, which captures the probability of extreme vehicle-pedestrian conflicts leading to a crash-equivalent condition. The risk of crash is expressed as in Equation (7).

$$RC_i = Pr(z_i > 0) = 1 - G_i(0)$$

$$= \begin{pmatrix} 1 - exp\left(-\left[1 - \xi_i \frac{\mu_i}{\sigma_i}\right]^{-\frac{1}{\xi}}\right), \; for \; \xi \neq 0 \\ 1 - exp\left[-exp\left(\frac{\mu_i}{\sigma_i}\right)\right], \; for \; \xi = 0 \end{pmatrix} \tag{7}$$

Here, $RC_i$ represents the crash risk in signal cycle $i$, where a value of zero indicates a safe cycle and a positive value indicates a potential crash occurrence within that cycle.

## *2.7. Modified Crash Risk Formulation*

Equation (7) has been widely adopted to model crash probability [8,10,12]. However, this direct estimation assumes that all risk-taking events are equally hazardous. In highly heterogeneous and congested traffic environments such as Dhaka, this assumption may not hold. Numerous studies have shown that risk-taking behaviour is deeply embedded in everyday pedestrian movement, making high-risk interactions a common and often necessary practice rather than an exception.

Pedestrians in Dhaka often accept very short gaps (≤2 s), run while crossing, or use mid-block and conflict-zone crossings instead of designated facilities [27]. These behaviours are linked to user characteristics, narrow or discontinuous medians, uncontrolled intersections, and high traffic volumes. They are also reinforced by the inconvenience of formal facilities such as footbridges and overpasses [28] and by preferences for short walking distances [29]. As a result, most pedestrians cross at grade even when traffic speeds and flows are high.

Risk Homeostasis Theory [34,35] suggests that individuals maintain a target level of perceived risk by adjusting their behaviour. These behavioural and infrastructural realities create a persistently elevated baseline of risk, where pedestrians continuously engage in actions that would be considered unsafe in more regulated environments. Environmental constraints, including blocked or damaged footpaths, poorly placed crosswalks, lack of median guard rails, and inadequate lighting, further normalize these behaviours [32,33]. Under these conditions, the conventional crash risk ($RC_i$) remains high in most cycles and cannot differentiate unusually dangerous events from typical behaviour.

To address this, a Modified Crash Risk (MRC) metric is proposed as a calibration strategy that adjusts the crash risk for each signal cycle relative to the site's habitual risk level. Risk Homeostasis Theory motivates this metric, but this study does not test the theory. The formulation is given in Equation (8).

$$MRC_i = \begin{cases} RC_i - \overline{RC} - Z_{cr} * \sigma_{RC}, & \overline{RC} \gg 0: \\ RC_i, & otherwise \end{cases} \tag{8}$$

Here, $MRC_i$ is the non-negative modified crash risk in signal cycle $i$. $\overline{RC}$ and $\sigma_{RC}$ denote the average and standard deviation of the crash risk for each cycle, respectively. $Z_{cr}$ indicates the z-critical value for a particular confidence interval. This formulation ensures that $MRC_i$ is non-negative and emphasizes only those cycles where the risk level clearly exceeds the site's normal exposure conditions.

Conceptually, this modification extends the exceedance-over-threshold principle to a behavioural setting, where the baseline ($\overline{RC} + Z_{cr} * \sigma_{RC}$) represents the routine level of risk at each site. By adjusting for this baseline, MRC is designed to identify cycles that represent breakdowns in pedestrian-vehicle interactions, such as chaotic crossings or sudden increases in pedestrian flow. It is therefore intended as a more context-sensitive measure of crash risk for Dhaka's mixed traffic.

## 3. Data collection and pre-processing

### *3.1. Data collection*

Intersections were selected using police-reported crash data from 2016–2020 obtained from the Accident Research Institute (ARI). To ensure a representative sample of Dhaka's urban network, nine intersections were chosen: four three-legged sites (Abul Hotel, Bonolota Market, Mirpur 1 Bus Stop, Shishu Mela) and five four-legged sites (Motsho Bhaban, Paltan, Shahbagh, Bijoy Sarani, Banglamotor). None of these locations have a dedicated pedestrian phase. Pedestrians cross individually during low traffic but form groups when traffic flow increases. Conflict data were recorded using two cameras at each site, installed at heights of 6–10 m to cover all movements. Videos were collected for two hours during weekday peak periods under favourable weather. Weather and time of day were intentionally restricted to keep conditions comparable across sites. The dataset therefore covers 18 hours of weekday peak period observation at nine intersections in one city. Table A1 (Appendix) summarizes the data collection details for each site, and Figure A2 illustrates camera coverage.

Conflicts were identified by analysing overlaps between predicted pedestrian and vehicle paths. A PET value near zero indicates a narrowly avoided collision. PET thresholds vary across studies: some studies used PET $\leq 6$ s [8,10], whereas some other studies applied PET $\leq 4$ s [44,45] to capture more critical interactions. Zheng et al. [46] showed that moderate variations in threshold selection have minimal impact on model estimation as long as the threshold remains well above the extreme boundary. Based on these considerations, this study defined a vehicle-pedestrian conflict when PET $< 5$ s. This value also reflects Dhaka's congested, non-lane-based traffic, where vehicles often move slowly. For the same distance between a vehicle and a pedestrian, PET is longer at low speed than at high speed. A 5 s threshold therefore keeps close interactions that a lower threshold could miss.

### *3.2. Signal cycle-level covariates*

An automated covariate extraction algorithm was developed to manage the multidimensional data required at the signal-cycle level. The algorithm synchronizes datasets by timestamp, assigns each trajectory and conflict to its corresponding cycle, and incorporates signal-phase information. It then identifies cycles involving pedestrian conflicts with different vehicle types and extracts key variables such as flow and speed for all road users, along with their conflicting flow and speed. Although this research focuses on vehicle-pedestrian interactions, the algorithm is adaptable to other crash types, including rear-end collisions.

Road-user counts were obtained for each cycle at all intersections. Due to Dhaka's mixed traffic, vehicle counts were converted to Passenger Car Units (PCU) following the Revised Strategic Transport Plan [47]. Road users were grouped into pedestrians, MVs, and NMVs, as these categories differ in speed, protection level, and associated crash severity [7,48]. Conflicting flow was defined as the flow of road users involved in conflict events, which was substantial across sites due to heterogeneous movement patterns and limited enforcement.

Speeds were derived from object coordinates and timestamps after camera calibration, which converted pixel measurements to real-world distances. Space-mean speeds were then computed for each cycle. Before inclusion in the extreme value model, a correlation analysis was performed to examine relationships among variables. Only covariates without strong mutual correlations were retained (Figure A1, Appendix). The selected variables include flow of MVs and pedestrians, speed of MVs and pedestrians, conflicting flow and conflicting speed of MVs, and conflicting speed of NMVs. Most of these variables showed weak to moderate correlations (with one pair reaching −0.71), which was considered acceptable for model estimation. Variables with high correlation, such as NMV flow, NMV speed, conflicting NMV flow, conflicting pedestrian flow and speed, were excluded during modelling unless statistically significant. Covariate selection was therefore partly data driven. The selected covariates should be interpreted as a practical, sample-specific set of predictors rather than a definitive list of factors influencing pedestrian crash risk.

Crash data were collected to benchmark the model, as summarized in Table 2. The dataset includes cycle counts, crash counts, and covariates for each intersection. The proposed framework was applied to daytime vehicle-pedestrian crashes extracted from video data, and the selected intersections recorded thirty-six such crashes between 2016 and 2020.

# 4. Model development and performance evaluation

## *4.1. Model specification and estimation*

Four types of Bayesian extreme value models were estimated by incorporating covariates into the GEV distribution through linear (Models 1 and 2) and non-linear (Models 3 and 4) link functions. Model 1 is the stationary baseline ($\alpha_{\mu 1} = \alpha_{\phi 1} = 0$), where no covariates are added to the location or scale parameters. Model 2 applies a linear link to the location parameter and a log-linear link to the scale parameter, resulting in three variants: Model 2(a) includes covariates only in the location parameter, Model 2(b) only in the scale parameter, and Model 2(c) in both.

Model 3 introduces non-linear link functions for the location parameter while maintaining a log-linear structure for the scale parameter. Two variants were estimated: Model 3(a) adds covariates only to the location parameter, and Model 3(b) adds covariates to both the location and scale parameters. Model 4 represents the most complex structure, applying non-linear link functions to both parameters to explore fully non-stationary relationships.

Additional configurations, such as incorporating non-linear link functions for the scale parameter while retaining linear functions for the location parameter, and those adding covariates exclusively to either the scale or location parameters in Models 3 and 4, were tested but did not converge. The shape parameter was not parameterized due to convergence challenges and imprecise estimation [39]. Several covariate combinations were tested, and model selection was based on DIC.

The dependent variable, PET, was used as the key traffic conflict indicator for each signal cycle in all models. Each of the seven sub-models was estimated using two independent MCMC chains with different initial values. A total of 76,000 iterations were conducted, with the first 26,000 removed as burn-in and the remaining 50,000 used to generate posterior estimates. Convergence was evaluated through trace plots and Gelman-Rubin statistics. For Model 3(a), the trace plots showed well-mixed chains and most Gelman-Rubin statistics were below 1.1, indicating stable estimation [10]. Figures A4 and A5 (Appendix) display these diagnostics for Model 3(a), where covariates were added only to the location parameter.

Because vehicle and pedestrian speeds are used to compute PET, directly including instantaneous speeds in the model could create endogeneity. To avoid this, aggregated space-mean speeds at the cycle level were used as covariates instead.

Table 2. Summary of traffic flow, speed metrics, conflict incidents, and crash data

| **Parameter** | | **$F_{MV}$ (PCU)** | **$F_P$** | **$S_{MV}$ (m/s per PCU)** | **$S_P$ (m/s)** | **$CF_{MV}$ (PCU)** | **$CS_{MV}$ (m/s per PCU)** | **$CS_{NMV}$ (m/s per PCU)** | **PET (s)** | **Cycle timings (min)** | **Number of cycles** | **Total conflicts** | **Crash records (2016-20)** |
|---|---|---|---|---|---|---|---|---|---|---|---|---|---|
| **Abul Hotel** | Mean | 824.79 | 1477.61 | 1.91 | 0.70 | 337.08 | 2.21 | 0.67 | 2.39 | 6.67 | 18 | 29744 | 4 |
| | S. D. | 50.53 | 145.87 | 0.16 | 0.03 | 35.37 | 0.17 | 0.06 | 0.07 | | | | |
| | Max | 901.75 | 1737 | 2.29 | 0.75 | 397.75 | 2.45 | 0.77 | 2.49 | | | | |
| | Min | 748.75 | 1279 | 1.66 | 0.62 | 279.75 | 1.89 | 0.59 | 2.20 | | | | |
| **Bonolota Market** | Mean | 503.08 | 199.22 | 4.01 | 1.87 | 123.01 | 4.10 | 1.99 | 2.22 | 3.33 | 36 | 8750 | 3 |
| | S. D. | 33.18 | 33.73 | 0.58 | 0.24 | 27.93 | 0.63 | 0.47 | 0.15 | | | | |
| | Max | 568.50 | 251 | 5.34 | 2.25 | 179.75 | 5.63 | 3.07 | 2.57 | | | | |
| | Min | 430.75 | 104 | 2.43 | 1.02 | 64.00 | 2.75 | 1.25 | 1.92 | | | | |
| **Mirpur 1 Bus Stop** | Mean | 343.69 | 1165.92 | 3.09 | 1.70 | 90.40 | 4.63 | 1.50 | 2.39 | 5.00 | 24 | 27568 | 4 |
| | S. D. | 59.08 | 106.08 | 0.74 | 0.30 | 16.89 | 1.30 | 0.26 | 0.06 | | | | |
| | Max | 456.25 | 1365 | 4.95 | 2.21 | 127 | 7.42 | 2.01 | 2.50 | | | | |
| | Min | 221.50 | 996 | 2.05 | 0.98 | 61.75 | 2.69 | 1.07 | 2.18 | | | | |
| **Shishu Mela** | Mean | 909.58 | 450 | 3.53 | 1.65 | 139.15 | 5.31 | 1.83 | 2.37 | 5.00 | 24 | 7599 | 2 |
| | S. D. | 86.89 | 179.25 | 0.61 | 0.74 | 23.56 | 1.26 | 0.74 | 0.12 | | | | |
| | Max | 1043.75 | 1102 | 5.27 | 2.32 | 197 | 8.05 | 3.53 | 2.59 | | | | |
| | Min | 699.50 | 265 | 2.63 | 0.85 | 100 | 3.50 | 0.63 | 2.15 | | | | |

Abbreviations: $F_{MV}$ = Flow of motorized vehicles; $F_P$ = Flow of pedestrians; $S_{MV}$ = Speed of motorized vehicles; $S_P$ = Speed of pedestrians; $CF_{MV}$ = Conflicting flow of motorized vehicles; $CS_{MV}$ = Conflicting speed of motorized vehicles; $CS_{NMV}$ = Conflicting speed of non-motorized vehicles; PCU = Passenger Car Unit; S.D. = Standard deviation.

| Parameter | | $F_{MV}$ (PCU) | $F_P$ | $S_{MV}$ (m/s per PCU) | $S_P$ (m/s) | $CF_{MV}$ (PCU) | $CS_{MV}$ (m/s per PCU) | $CS_{NMV}$ (m/s per PCU) | PET (s) | Cycle timings (min) | Number of cycles | Total conflicts | Crash records (2016-20) |
|---|---|---|---|---|---|---|---|---|---|---|---|---|---|
| **Motsho Bhaban** | Mean | 1639.33 | 248.95 | 3.59 | 1.73 | 187.99 | 4.41 | 2.00 | 2.36 | 6.00 | 20 | 8662 | 8 |
| | S. D. | 133.88 | 50.52 | 0.33 | 0.24 | 33.66 | 0.44 | 0.30 | 0.11 | | | | |
| | Max | 1818.25 | 350 | 4.09 | 2.26 | 245.50 | 5.14 | 3.04 | 2.61 | | | | |
| | Min | 1238.25 | 143 | 2.98 | 1.30 | 136.50 | 3.75 | 1.61 | 2.19 | | | | |
| **Paltan** | Mean | 915.84 | 198.29 | 6.61 | 1.96 | 107.39 | 8.56 | 3.26 | 2.47 | 5.00 | 24 | 16229 | 4 |
| | S. D. | 144.78 | 30.05 | 1.17 | 0.55 | 19.50 | 1.58 | 0.58 | 0.06 | | | | |
| | Max | 1190 | 261 | 8.57 | 2.35 | 151.75 | 10.60 | 4.31 | 2.63 | | | | |
| | Min | 685 | 134 | 5.01 | 1.32 | 68 | 6.58 | 1.94 | 2.38 | | | | |
| **Shahbagh** | Mean | 553.70 | 192.93 | 5.17 | 1.82 | 95.34 | 7.37 | 2.03 | 2.53 | 4.00 | 30 | 11342 | 3 |
| | S. D. | 120 | 37.74 | 1.10 | 0.89 | 39.24 | 1.23 | 0.95 | 0.08 | | | | |
| | Max | 855.25 | 317 | 7.80 | 2.22 | 198 | 9.57 | 3.11 | 2.71 | | | | |
| | Min | 339.75 | 142 | 4.31 | 1.05 | 43 | 5.90 | 1.37 | 2.35 | | | | |
| **Bijoy Sarani** | Mean | 923.49 | 181.13 | 6 | 1.91 | 109.78 | 8.06 | 2.14 | 2.22 | 4.00 | 30 | 5865 | 4 |
| | S. D. | 112.25 | 40.01 | 0.93 | 0.45 | 36.98 | 1.17 | 0.63 | 0.25 | | | | |
| | Max | 1163.50 | 281 | 7.21 | 2.39 | 173 | 9.96 | 3.12 | 2.82 | | | | |
| | Min | 698.75 | 104 | 4.44 | 1.18 | 46 | 5.81 | 1.55 | 1.76 | | | | |
| **Banglamoto r** | Mean | 590.52 | 182.94 | 4.94 | 1.79 | 175.54 | 6.81 | 2.35 | 2.41 | 3.75 | 32 | 25223 | 4 |
| | S. D. | 178.77 | 46.58 | 1.23 | 0.74 | 23.61 | 1.27 | 0.81 | 0.06 | | | | |
| | Max | 891.50 | 284 | 6.18 | 2.48 | 228 | 9.10 | 4.04 | 2.54 | | | | |
| | Min | 361.75 | 122 | 3.98 | 1.15 | 138.50 | 5.12 | 1.39 | 2.28 | | | | |

Abbreviations: FMV = Flow of motorized vehicles; FP = Flow of pedestrians; SMV = Speed of motorized vehicles; SP = Speed of pedestrians; CFMV = Conflicting flow of motorized vehicles; CSMV = Conflicting speed of motorized vehicles; CSNMV = Conflicting speed of non-motorized vehicles; PCU = Passenger Car Unit; S.D. = Standard deviation.

Table 3. Summary of the model estimation results

| Model | | Model 1 | | | | Model 2(a) | | | | Model 2(b) | | | | Model 2(c) | | | |
|---|---|---|---|---|---|---|---|---|---|---|---|---|---|---|---|---|---|
| Parameter | | Mean | S. D. | 2.5% | 97.5% | Mean | S. D. | 2.5% | 97.5% | Mean | S. D. | 2.5% | 97.5% | Mean | S. D. | 2.5% | 97.5% |
| Location | $\mu_0$ | 2.319 | 0.011 | 2.295 | 2.34 | 2.385 | 0.029 | 2.328 | 2.423 | 2.099 | 0.0003 | 2.098 | 2.099 | 2.153 | 0.058 | 2.007 | 2.244 |
| | $\mu_{FMV}$ | - | - | - | - | -0.0004 | 0.0002 | -0.001 | -0.0001 | - | - | - | - | 0.0002 | 0.0002 | -0.001 | 0.001 |
| | $\mu_{FP}$ | - | - | - | - | -0.0001 | 0.0001 | -0.0003 | 0.0002 | - | - | - | - | 0.0004 | 0.0004 | -0.001 | 0.001 |
| | $\mu_{SMV}$ | - | - | - | - | -0.034 | 0.006 | -0.044 | -0.021 | - | - | - | - | -0.009 | 0.008 | -0.025 | 0.004 |
| | $\mu_{SP}$ | - | - | - | - | 0.012 | 0.004 | 0.005 | 0.02 | - | - | - | - | 0.013 | 0.005 | 0.002 | 0.022 |
| | $\mu_{CFMV}$ | - | - | - | - | 0.001 | 0.001 | -0.001 | 0.003 | - | - | - | - | 0.0004 | 0.002 | -0.006 | 0.003 |
| | $\mu_{CSMV}$ | - | - | - | - | 0.009 | 0.003 | 0.002 | 0.015 | - | - | - | - | 0.003 | 0.004 | -0.003 | 0.011 |
| | $\mu_{CSNMV}$ | - | - | - | - | 0.01 | 0.003 | 0.002 | 0.016 | - | - | - | - | 0.005 | 0.005 | -0.003 | 0.015 |
| Scale | $\sigma_0$ | 1.36 | 0.002 | 1.355 | 1.364 | 1.359 | 0.007 | 1.347 | 1.364 | 12.01 | 0.0004 | 12.01 | 12.01 | 2.389 | 2.67 | 1.337 | 9.736 |
| | $\sigma_{FMV}$ | - | - | - | - | - | - | - | - | -0.003 | 0.0002 | -0.004 | -0.003 | -0.004 | 0.01 | -0.031 | 0.0003 |
| | $\sigma_{FP}$ | - | - | - | - | - | - | - | - | -0.011 | 0.0002 | -0.011 | -0.011 | -0.005 | 0.013 | -0.048 | 0.001 |
| | $\sigma_{SMV}$ | - | - | - | - | - | - | - | - | 0.00001 | 0.0003 | -0.0002 | 0.0004 | -0.006 | 0.018 | -0.055 | 0.004 |
| | $\sigma_{SP}$ | - | - | - | - | - | - | - | - | -0.007 | 0.0004 | -0.008 | -0.007 | -0.003 | 0.01 | -0.033 | 0.004 |
| | $\sigma_{CFMV}$ | - | - | - | - | - | - | - | - | -0.018 | 0.0003 | -0.018 | -0.017 | -0.005 | 0.01 | -0.033 | -0.0001 |
| | $\sigma_{CSMV}$ | - | - | - | - | - | - | - | - | -0.005 | 0.0003 | -0.006 | -0.005 | -0.006 | 0.013 | -0.047 | 0.001 |
| | $\sigma_{CSNMV}$ | - | - | - | - | - | - | - | - | -0.005 | 0.0001 | -0.005 | -0.005 | -0.001 | 0.004 | -0.013 | 0.001 |
| Shape | $\xi_0$ | -0.41 | 0.013 | -0.433 | -0.381 | -0.413 | 0.013 | -0.437 | -0.385 | -3.87 | 0.004 | -3.874 | -3.864 | -0.553 | 0.398 | -1.747 | -0.379 |
| DIC | | 7164 | | | | 7125 | | | | 16720 | | | | 7087 | | | |

Abbreviations: FMV = Flow of motorized vehicles; FP = Flow of pedestrians; SMV = Speed of motorized vehicles; SP = Speed of pedestrians; CFMV = Conflicting flow of motorized vehicles; CSMV = Conflicting speed of motorized vehicles; CSNMV = Conflicting speed of non-motorized vehicles; DIC = Deviance Information Criterion; S.D. = Standard deviation.

| Model | | Model 3(a) | | | | | Model 3(b) | | | | | Model 4 | | | | |
|---|---|---|---|---|---|---|---|---|---|---|---|---|---|---|---|---|
| Parameter | | Mean | S. D. | 2.5% | 97.5% | θ | Mean | S. D. | 2.5% | 97.5% | θ | Mean | S. D. | 2.5% | 97.5% | θ |
| Location | $\mu_0$ | 1.759 | 0.705 | 0.044 | 2.831 | - | 2.057 | 0.406 | 1.165 | 3.002 | - | 1.952 | 0.044 | 1.871 | 2.046 | - |
| | $\mu_{FMV}$ | 0.513 | 0.835 | -0.793 | 2.141 | -0.591 | -28.81 | 77.65 | -247.3 | 86.11 | -1.011 | -4.66 | 6.515 | -14.44 | 0.816 | -0.099 |
| | $\mu_{FP}$ | 0.165 | 0.281 | -0.537 | 0.812 | -0.698 | -3.491 | 6.18 | -16.34 | -0.29 | -0.782 | -9.901 | 8.145 | -19.36 | 0.412 | -0.908 |
| | $\mu_{SMV}$ | 0.261 | 0.447 | -0.707 | 1.126 | -0.185 | 0.074 | 0.34 | -0.297 | 1.065 | 0.3 | -0.064 | 0.076 | -0.185 | 0.008 | -0.35 |
| | $\mu_{SP}$ | -0.12 | 0.473 | -1.063 | 0.561 | -1.027 | 0.333 | 0.498 | 0.003 | 1.808 | -0.448 | -0.084 | 0.047 | -0.168 | -0.017 | -0.157 |
| | $\mu_{CFMV}$ | -0.449 | 0.404 | -1.198 | 0.496 | 0.145 | -1.025 | 1.844 | -4.517 | 3.698 | -0.622 | -0.84 | 0.85 | -1.832 | -0.006 | -1.253 |
| | $\mu_{CSMV}$ | -0.14 | 0.396 | -1.227 | 0.545 | -0.97 | -0.078 | 0.363 | -0.617 | 0.679 | -0.964 | -0.764 | 0.605 | -1.605 | -0.152 | -0.078 |
| | $\mu_{CSNMV}$ | 0.381 | 0.597 | -0.132 | 1.978 | 0.14 | 0.28 | 0.438 | -0.025 | 1.545 | 0.909 | 0.008 | 0.003 | 0.0009 | 0.015 | 0.005 |
| Scale | $\sigma_0$ | 1.359 | 0.003 | 1.355 | 1.364 | - | 4.892 | 3.418 | 1.339 | 8.336 | - | 10.67 | 0.747 | 9.93 | 11.77 | - |
| | $\sigma_{FMV}$ | - | - | - | - | - | 0.00001 | 0.002 | -0.0002 | 0.0005 | 0.0003 | -0.039 | 0.01 | -0.049 | -0.026 | 0.932 |
| | $\sigma_{FP}$ | - | - | - | - | - | -0.009 | 0.01 | -0.019 | 0.0006 | 0.0014 | -0.025 | 0.015 | -0.042 | -0.01 | 0.736 |
| | $\sigma_{SMV}$ | - | - | - | - | - | -0.008 | 0.01 | -0.018 | 0.003 | -0.003 | -0.099 | 0.058 | -0.167 | -0.027 | -0.637 |
| | $\sigma_{SP}$ | - | - | - | - | - | -0.012 | 0.013 | -0.03 | 0.003 | 0.0004 | -0.037 | 0.012 | -0.059 | -0.024 | 0.633 |
| | $\sigma_{CFMV}$ | - | - | - | - | - | -0.009 | 0.009 | -0.017 | -0.0009 | -0.0008 | -0.03 | 0.019 | -0.056 | -0.009 | 0.951 |
| | $\sigma_{CSMV}$ | - | - | - | - | - | -0.01 | 0.009 | -0.022 | 0.0006 | 0.001 | -0.032 | 0.019 | -0.06 | -0.011 | 0.49 |
| | $\sigma_{CSNMV}$ | - | - | - | - | - | -0.001 | 0.002 | -0.007 | 0.0003 | 0.0011 | -0.046 | 0.064 | -0.206 | -0.007 | 0.817 |
| Shape | $\xi_0$ | -0.41 | 0.013 | -0.434 | -0.384 | - | -1.215 | 0.790 | -2.071 | -0.387 | - | -2.28 | 0.477 | -2.784 | -1.781 | - |
| DIC | | | | 7062 | | | | | 7064 | | | | | 15470 | | |

Abbreviations: FMV = Flow of motorized vehicles; FP = Flow of pedestrians; SMV = Speed of motorized vehicles; SP = Speed of pedestrians; CFMV = Conflicting flow of motorized vehicles; CSMV = Conflicting speed of motorized vehicles; CSNMV = Conflicting speed of non-motorized vehicles; DIC = Deviance Information Criterion; S.D. = Standard deviation; θ = exponent of the covariates of location or scale parameter.

### *4.2. Model comparison and selection*

Table 3 reports the DIC values for all seven sub-models. Models 1, 2(a), and 2(c) show acceptable performance, while Model 2(b) performs poorly, indicating that the scale parameter does not align well with covariates; it also shows the highest intercept value (12.01). Model 4 has a DIC of 15,470, showing a much poorer fit when non-linear forms are applied to both parameters. Among the seven models, Model 3 achieved the lowest DIC values, indicating that non-linear relationships in the location parameter combined with a log-linear relationship in the scale parameter provided the best fit to this dataset. Between the two variants, Model 3(a) has the lower DIC and the lower intercept estimates, while Model 3(b) carries very wide credible intervals. Model 3(a) is therefore selected as the final model.

The estimated Bayesian extreme value models were evaluated by comparing predicted crashes with crashes observed over a defined period. The mean number of estimated crashes over T years is computed using the formulation proposed by Zheng et al. [46], shown in Equation (9):

$$N = \frac{T}{t} * \sum_{i=1}^{m} MRC_i \tag{9}$$

Here, $N$ represents the expected number of crashes over duration T, where t is the video recording duration, $MRC_i$ denotes the modified crash risk (considering only positive MRC values), and m is the total number of cycles. In Equation (8), the parameter $Z_{cr}$ is assigned a value of 1.45, representing a 93% confidence level, which was chosen because it gave a close match between estimated and observed crashes for Model 3(a). Researchers applying this framework elsewhere should calibrate the confidence level to suit their conditions. Sensitivity analyses are recommended to identify a suitable confidence level for each setting, with the 93% level serving as a starting point rather than a fixed standard.

From 2016 to 2020, a total of 36 crashes were observed across the nine studied locations. In comparison, the mean estimated crashes for Models 1, 2(a), 2(b), 2(c), 3(a), 3(b), and 4 are 0, 137.47, 26.34, 697.64, 40.57, 566.25, and 78.81, respectively. Among these, Model 3(a) provides the closest mean estimate to the observed crashes, which is consistent with its selection by DIC. This closer agreement shows that the MRC formulation can reproduce the observed crash count for this dataset, although it does not show that the non-linear structure captures the true crash mechanism more accurately.

## 5. Discussion

### *5.1. Covariate estimates of the selected model*

Model 3(a) estimates how each covariate is associated with pedestrian crash risk at intersections. The coefficient terms in the location parameter indicate whether each covariate is positively or negatively associated with crash risk, while the exponent terms reflect the sensitivity of the relationship, with an exponent of zero indicating no influence. Together, these terms show how variations in traffic and pedestrian conditions may shape crash risk and can guide targeted safety improvements.

As shown in Table 3, pedestrian speed has the largest absolute exponent (−1.027) and a coefficient of −0.12. However, the 95% credible intervals of all covariate coefficients in Model 3(a) include zero, so the estimated effects should be read with caution. Table 4 summarizes the direction and sensitivity of all covariates included in the model, providing a concise overview of their roles in shaping crash risk at intersections. These interpretations are possible explanations of the estimated signs and exponents.

Table 4. Summary of covariate relationships with pedestrian crash risk at intersections

| Covariate | Estimated direction and possible reason | Possible interpretation of exponent |
|---|---|---|
| Flow of MVs | Positive: Raises crash risk due to higher pedestrian-vehicle interactions. | Large negative exponent: Effect diminishes at high volumes as congestion slows vehicles, creating stop-and-go conditions. |
| Flow of pedestrians | Positive: Increases exposure to potential crashes. | High negative exponent: Marginal risk increase diminishes at higher volumes as drivers become more cautious and vehicle speeds fall. |
| Speed of MVs | Positive: Higher speeds reduce available reaction time. | Relatively low negative exponent: Marginal effect of increasing speed diminishes at higher velocities as drivers adopt more cautious behaviours. |
| Speed of pedestrians | Negative: Initially reduces crash risk by minimizing time in conflict areas. | Largest negative exponent: Risk reduction diminishes at higher pedestrian speeds as focus on maintaining pace may reduce awareness. |
| Conflicting flow of MVs | Negative: Reduces crash risk due to slower speeds and cautious driving. | Smaller positive exponent: Marginal risk reduction diminishes as conflicting flow increases. |
| Conflicting speed of MVs | Negative: Initially reduces crash risk as drivers become more vigilant. | Strongly negative exponent: At very high speeds, reduced reaction time and longer stopping distances increase risk. |
| Conflicting speed of NMVs | Positive: Increases crash risk due to potential driver misjudgments and less protective structures. | Small positive exponent: While risk rises with speed, the rate of increase diminishes at higher speeds. |

### *5.2. Cycle-level risk patterns and practical implications*

For each signal cycle, a unique GEV distribution is generated under Model 3(a) using the covariates specific to that cycle. Figure A3 (Appendix) illustrates this by presenting three randomly selected cycles from each intersection, with positive crash-risk areas highlighted in red. Cycle-level assessment relies on the shape of the GEV distribution, where a positive crash risk occurs when the distribution's tail extends before PET reaches zero. All cycles across the studied intersections exhibit positive crash risk, which likely reflects Dhaka's heterogeneous traffic conditions, lack of lane discipline, and prevalent high-risk pedestrian behaviour.

Identifying cycles with elevated risk could help agencies apply timely interventions. Potential measures include allocating dedicated pedestrian green time, restricting permissive turning movements during pedestrian crossings, and using cycle-by-cycle crash-risk estimates to proactively manage safety in mixed-traffic environments.

## 6. Conclusions

This study develops a cycle-level vehicle-pedestrian crash risk framework for signalized intersections in mixed, non-lane-based traffic, explicitly accounting for vehicle heterogeneity by integrating MVs and NMVs within a unified modelling framework. Using the BM approach of EVT with a GEV distribution, extreme conflicts based on signal-cycle PET values are linked to crash risk through a hierarchical Bayesian structure at four three-legged and five four-legged intersections in Dhaka. The results suggest that incorporating traffic heterogeneity is important for realistic pedestrian safety assessment under complex urban conditions.

A key methodological contribution is the estimation and comparison of Bayesian GEV models with both linear and non-linear link functions. Allowing GEV parameters to vary with covariates enables the model to represent non-stationary and non-linear dynamics. Model 3, which uses a non-linear link function for the location parameter and a log-linear link function for the scale parameter, performs best among the estimated models, with Model 3(a) (covariates only in the location parameter) achieving the lowest DIC. These results suggest that non-linear link functions can improve model flexibility in representing complex, non-monotonic relationships between traffic variables and crash risk.

The study also proposes a behaviour-normalized MRC metric as a calibration strategy to reduce the overestimation produced by conventional formulations in environments where risk-taking is habitual. By adjusting crash risk relative to a local baseline, MRC is designed to better reflect conditions in Dhaka's congested mixed traffic. In addition, the posterior estimates indicate a negative association between pedestrian speed and crash risk, and positive associations for MV flow, MV speed, pedestrian flow, and the conflicting speed of NMVs. Because of the uncertainty in these estimates, the observed directions should be considered suggestive rather than conclusive.

This study should be viewed with a few points in mind. The data cover only nine intersections in one city, with each site observed for two hours during weekday peak periods under favourable weather conditions. The benchmark relies on 36 police-reported crashes, the MRC confidence level was tuned to this dataset, and not all tested specifications converged. The framework is therefore calibrated to a specific traffic setting, and its transferability remains to be evaluated. Nevertheless, the findings suggest that EVT-based crash risk modelling combined with non-linear link functions and behavioural normalization is a promising framework for cycle-level safety analysis in similar mixed-traffic conditions. The approach could support targeted interventions such as signal timing changes and turning-movement controls at high-risk cycles. Future work could validate the framework with larger datasets and independent locations, and incorporate socio-demographic factors such as age and gender.

## CRediT authorship contribution statement

**Parvez Anowar:** Writing – original draft, Visualization, Validation, Software, Methodology, Investigation, Data curation, Conceptualization**. Nazmul Haque:** Visualization, Validation, Software, Methodology, Investigation, Conceptualization**. Md Asif Raihan:** Writing – review & editing, Visualization, Validation, Methodology, Supervision, Conceptualization. **Md Hadiuzzaman:** Validation, Supervision, Project administration, Conceptualization.

## Funding statement

This research was supported by the Research Expenses (Regular) Fund (Docket No.: 1718, Sector No.: 3257103(A)) from the Department of Civil Engineering, Bangladesh University of Engineering and Technology (BUET), Bangladesh.

## Conflict of interest statement

The authors declare that they have no known competing financial interests or personal relationships that could have appeared to influence the work reported in this paper.

## Data availability statement

The data used in this study are available from the corresponding author upon reasonable request.

## ORCID iDs

Parvez Anowar: https://orcid.org/0009-0000-3656-237X
Nazmul Haque: https://orcid.org/0009-0006-4698-3366

Md Asif Raihan: https://orcid.org/0009-0005-1762-8528
Md Hadiuzzaman: https://orcid.org/0000-0002-5690-3351

## Appendix

Table A1. Timetable for data collection at study sites

| Intersection name | No. of camera | Date | Time | Duration |
|---|---|---|---|---|
| Abul Hotel | 2 | 11/12/2023 | 4 pm-6 pm | 2 h |
| Bonolota Market | 2 | 12/12/2023 | 4 pm-6 pm | 2 h |
| Mirpur 1 Bus Stop | 2 | 14/12/2023 | 4:30 pm-6:30 pm | 2 h |
| Shishu Mela | 2 | 17/12/2023 | 4 pm-6 pm | 2 h |
| Motsho Bhaban | 2 | 18/12/2023 | 4 pm-6 pm | 2 h |
| Paltan | 2 | 20/12/2023 | 4 pm-6 pm | 2 h |
| Shahbagh | 2 | 21/12/2023 | 4 pm-6 pm | 2 h |
| Bijoy Sarani | 2 | 27/12/2023 | 4:30 pm-6:30 pm | 2 h |
| Banglamotor | 2 | 28/12/2023 | 4:30 pm-6:30 pm | 2 h |

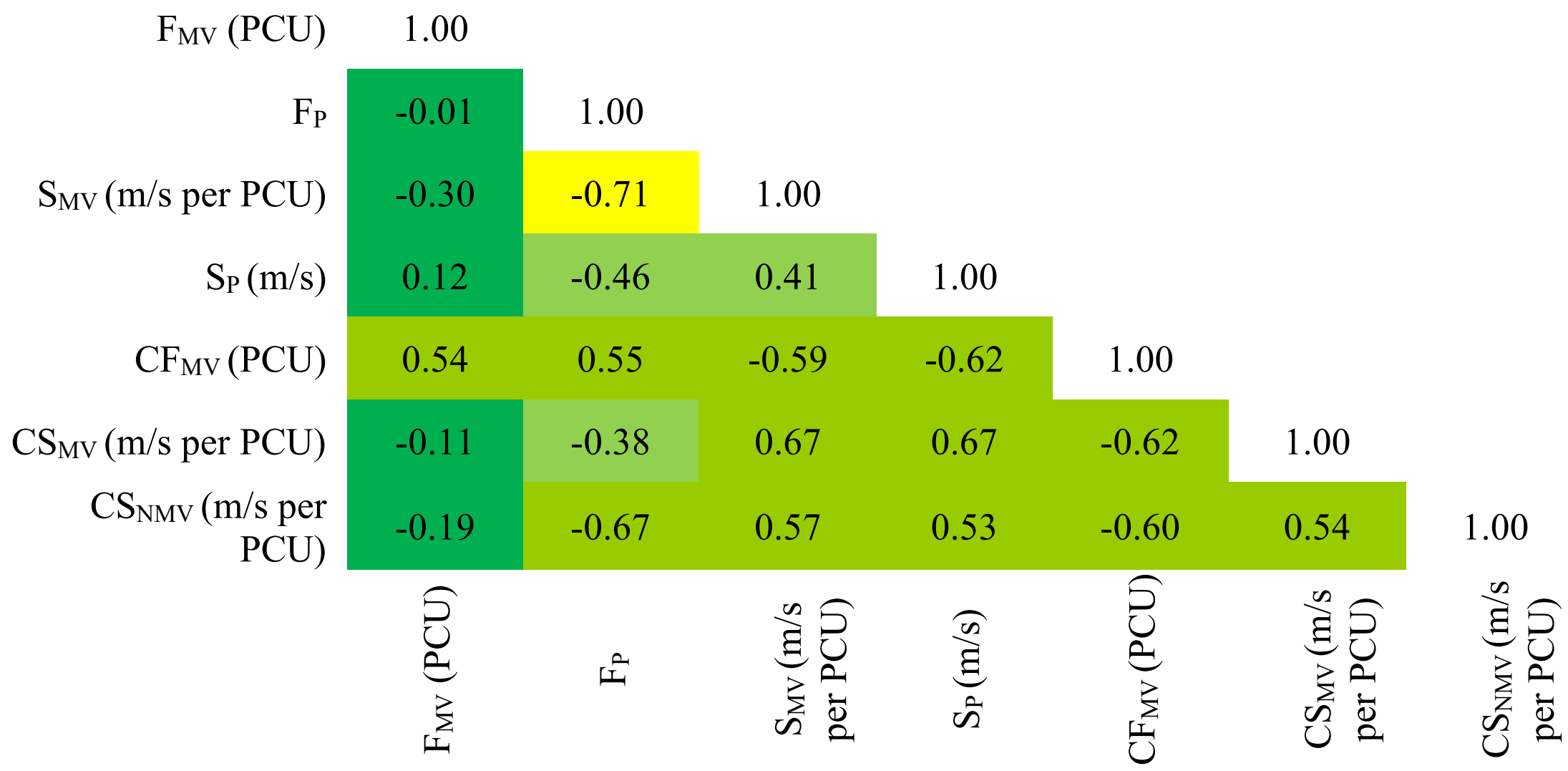

Figure A1. Correlation heatmap among covariates

Abbreviations: $F_{MV}$ = Flow of motorized vehicles; $F_P$ = Flow of pedestrians; $S_{MV}$ = Speed of motorized vehicles; $S_P$ = Speed of pedestrians; $CF_{MV}$ = Conflicting flow of motorized vehicles; $CS_{MV}$ = Conflicting speed of motorized vehicles; $CS_{NMV}$ = Conflicting speed of non-motorized vehicles; PCU = Passenger Car Unit

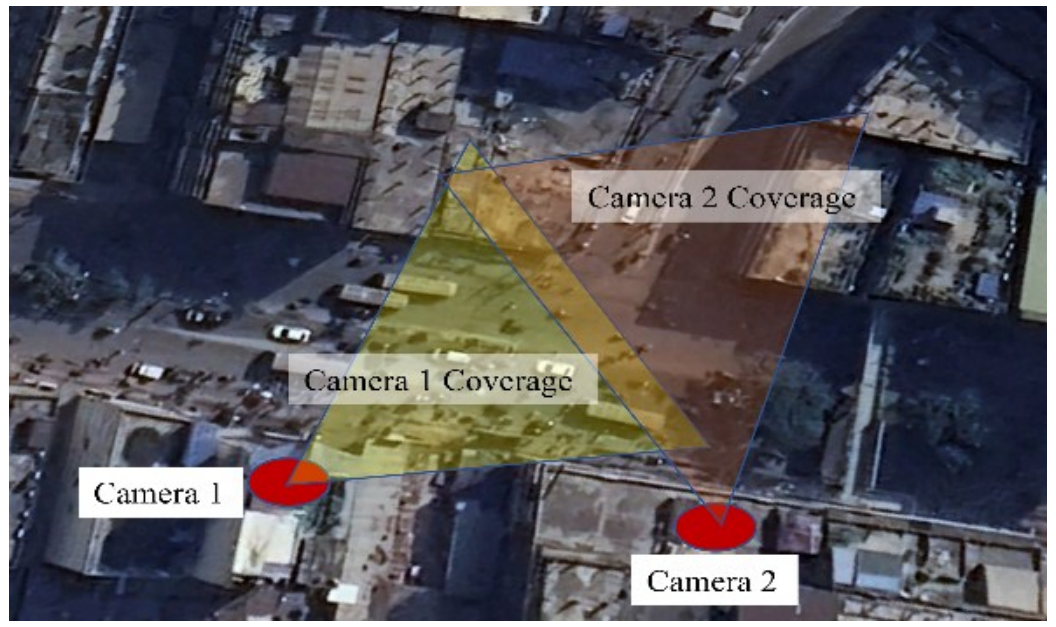

(a) Abul Hotel Intersection

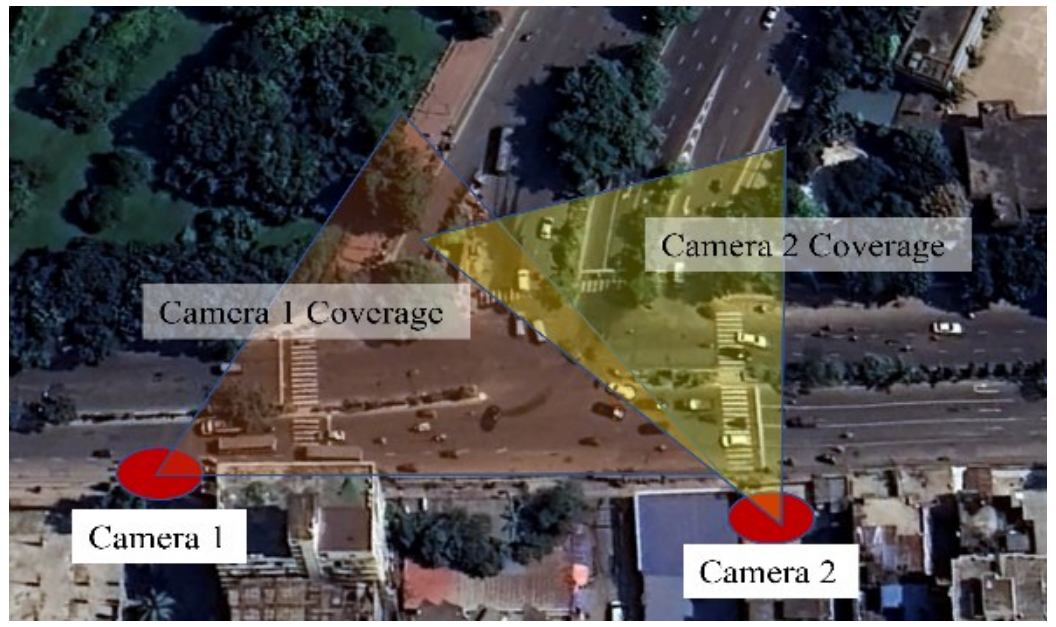

(b) Bonolota Market Intersection

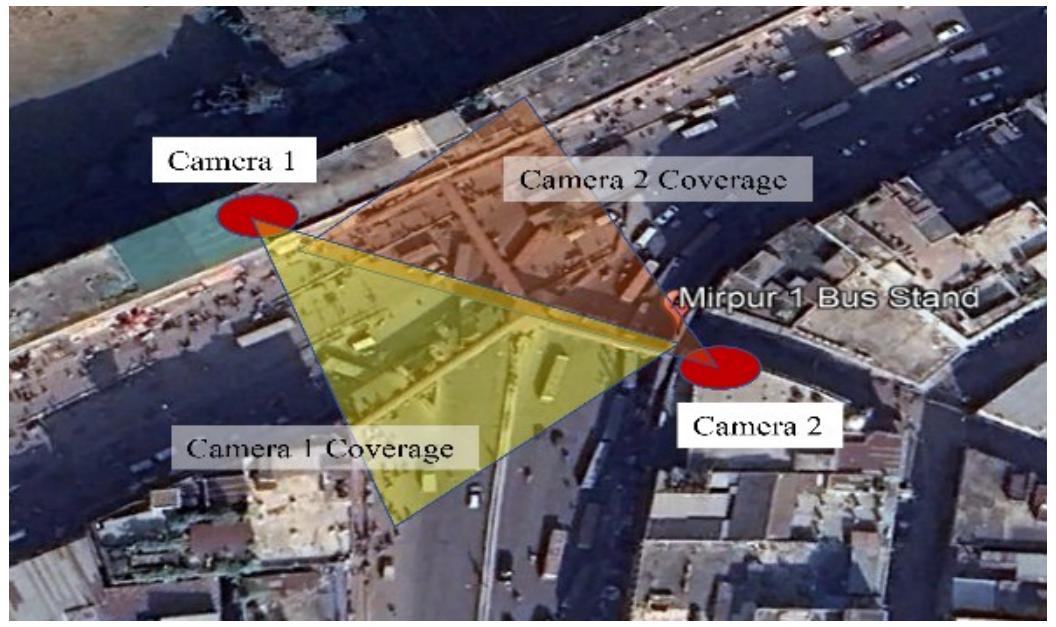

(c) Mirpur 1 Bus Stop Intersection

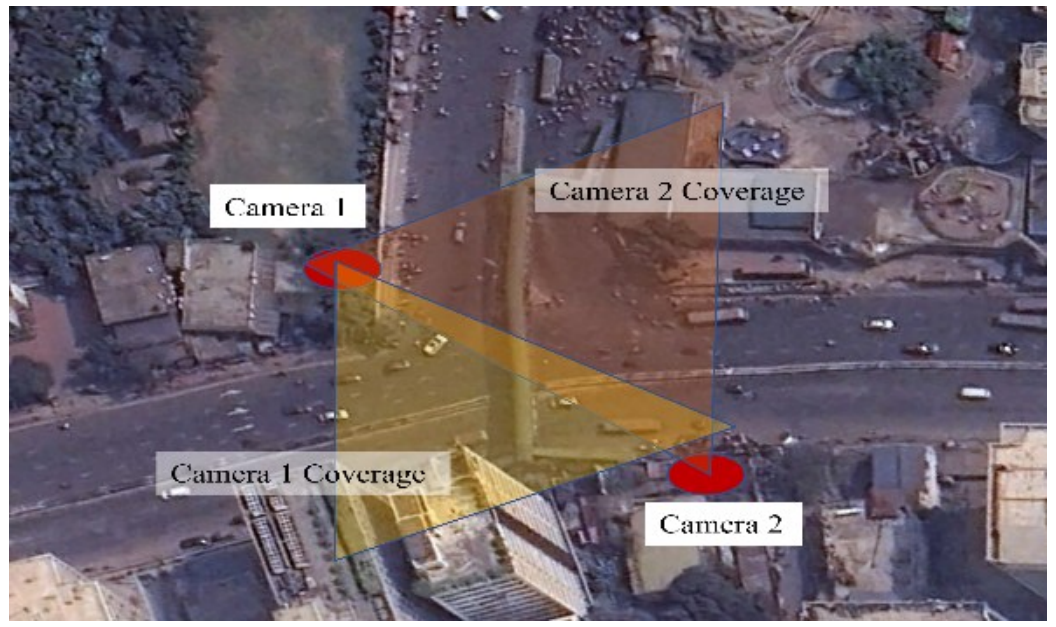

(d) Shishu Mela Intersection

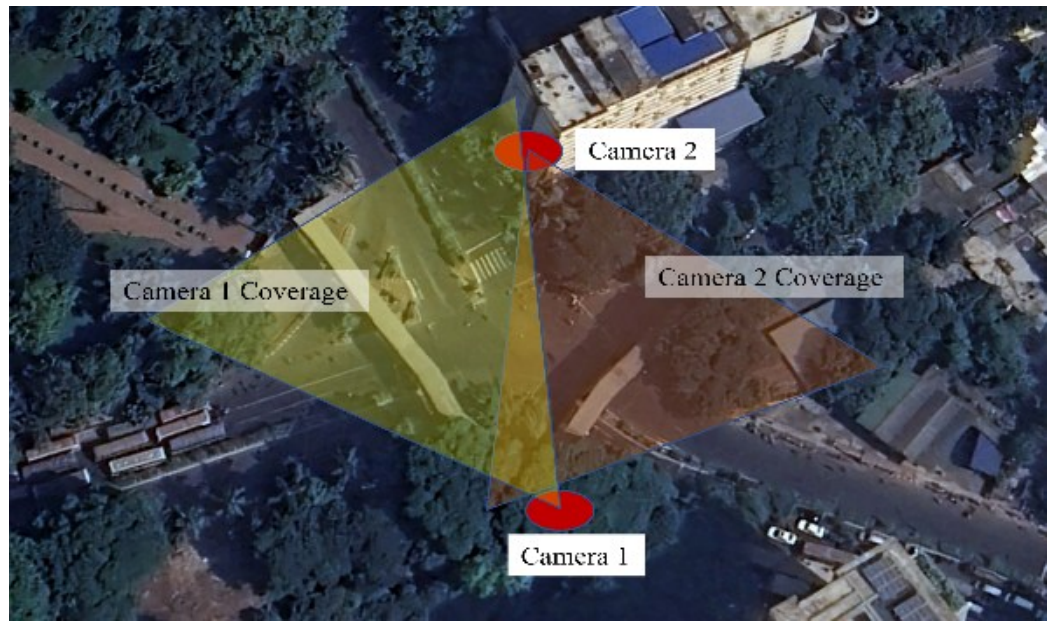

(e) Motsho Bhaban Intersection

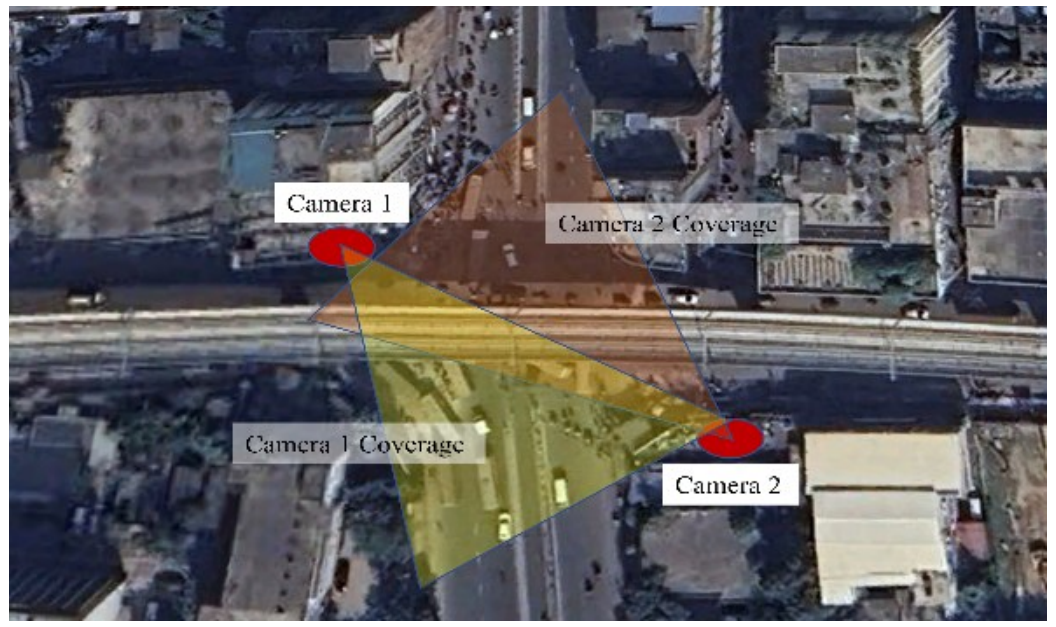

(f) Paltan Intersection

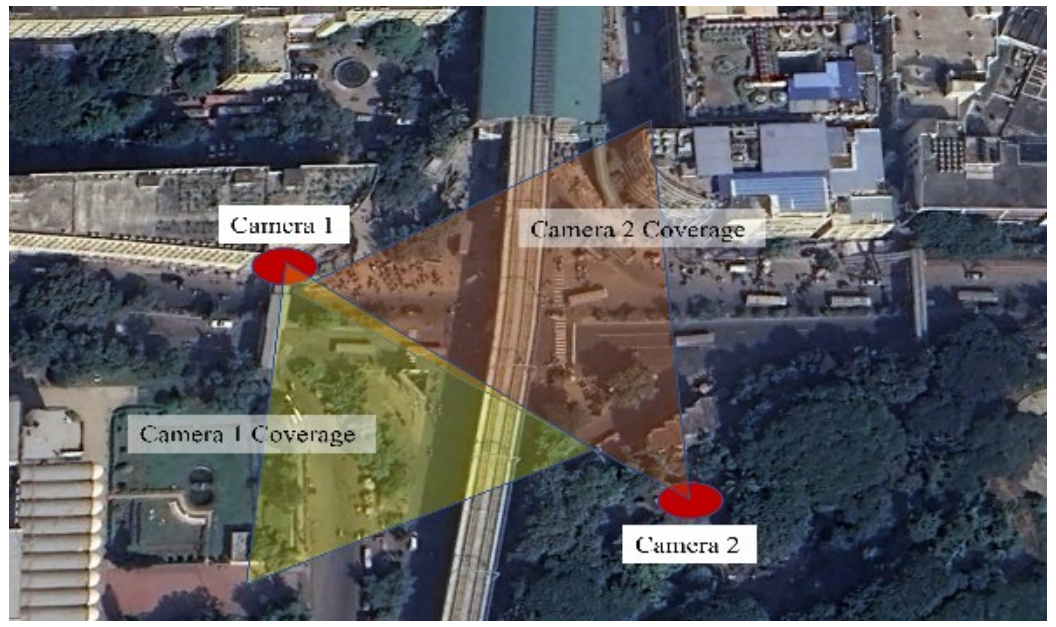

(g) Shahbagh Intersection

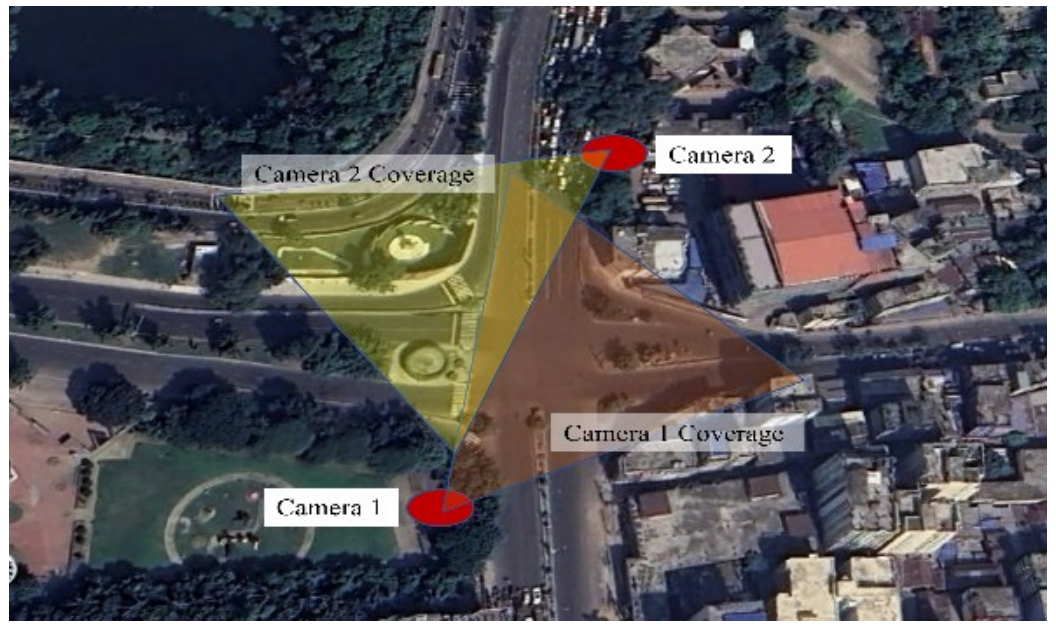

(h) Bijoy Sarani Intersection

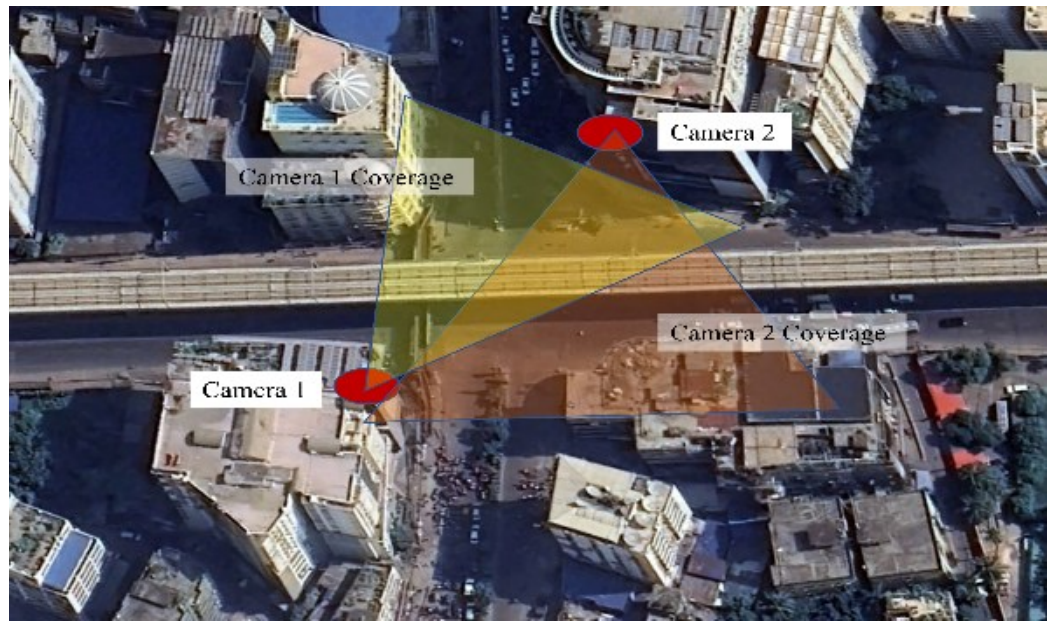

(i) Banglamotor Intersection

Figure A2. Camera positions at the study sites

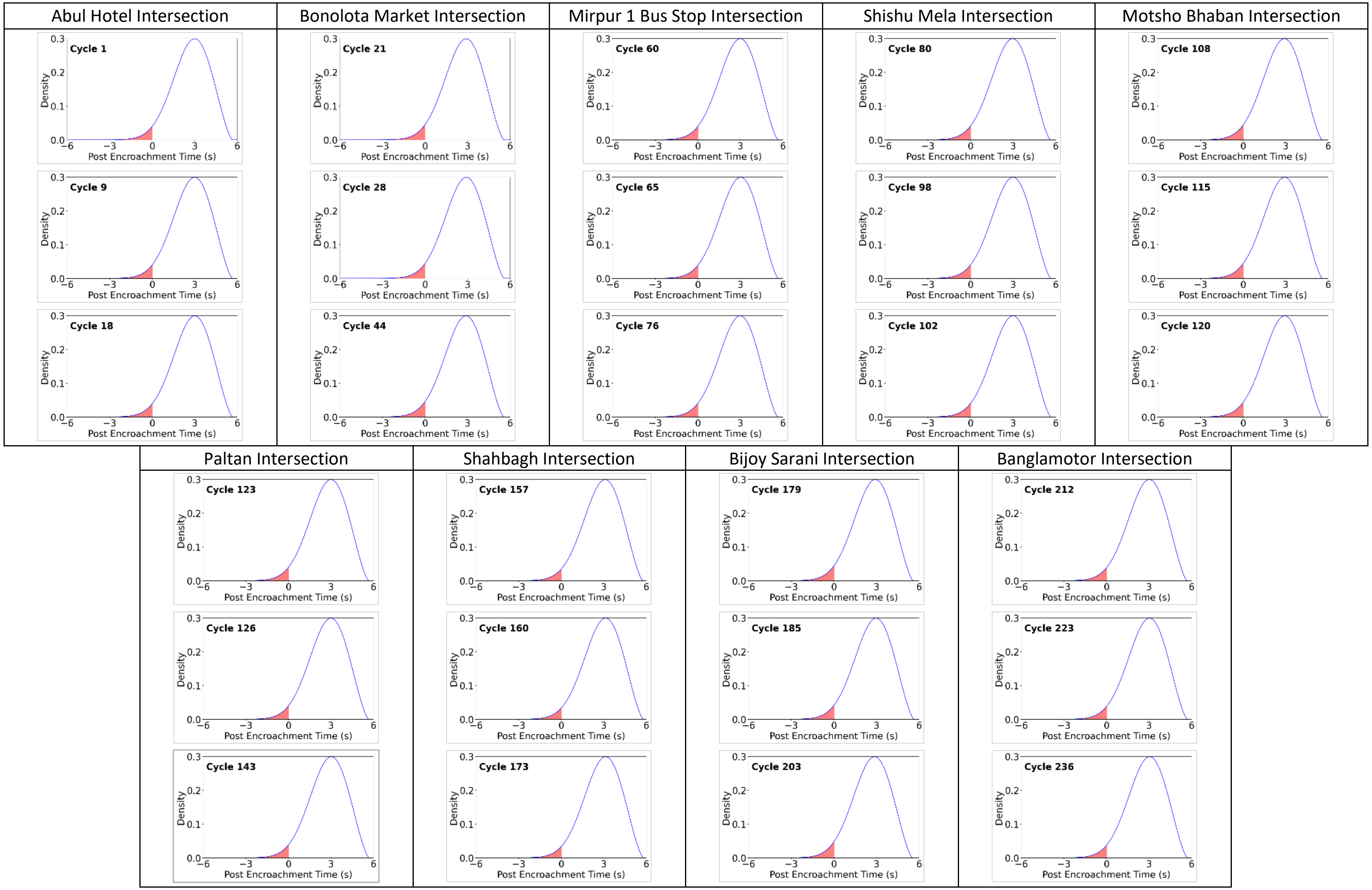

Figure A3. Estimated generalized extreme value distributions for sample cycles at each intersection using Model 3(a)

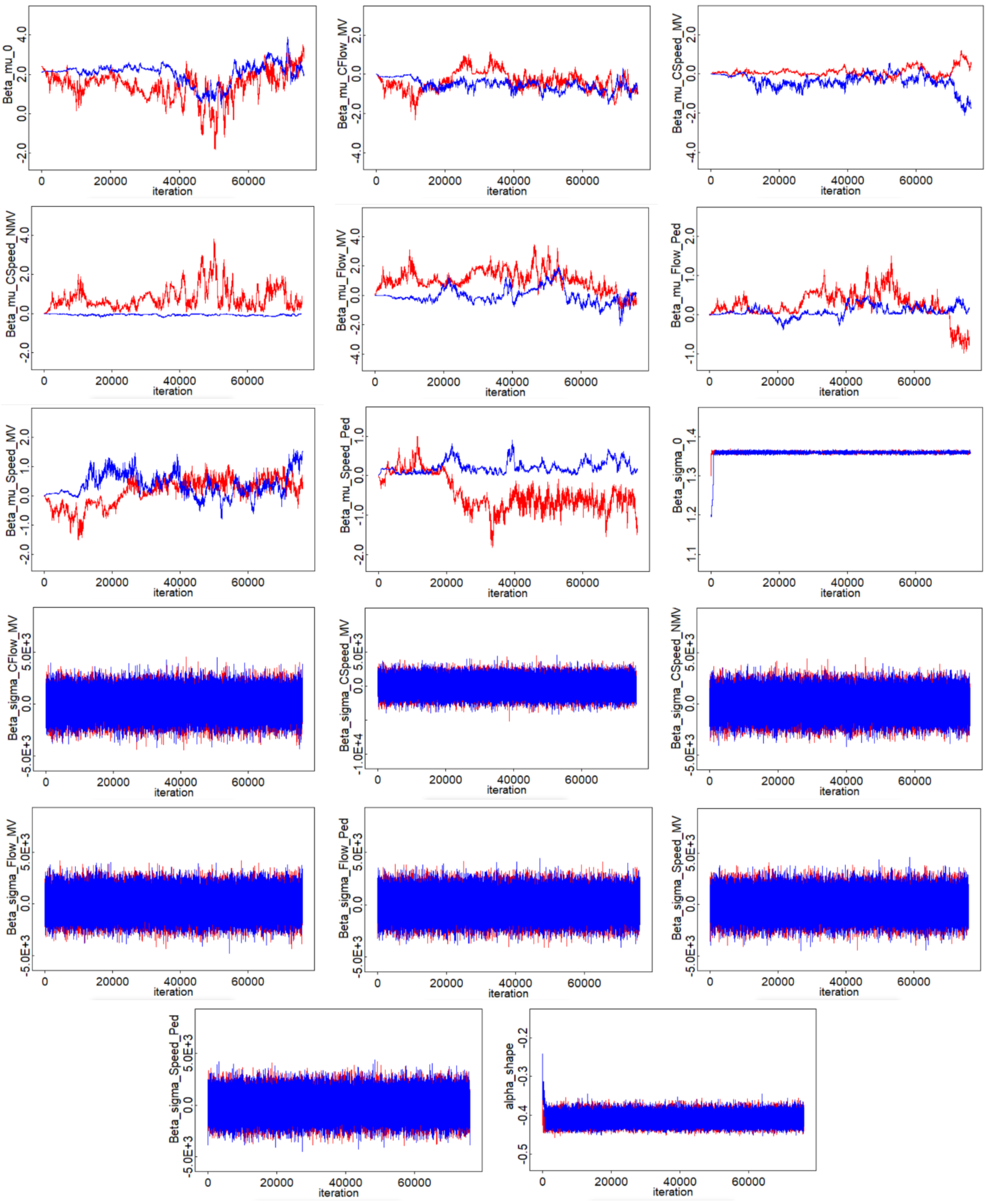

Figure A4. Trace plots of covariates of Model 3(a)

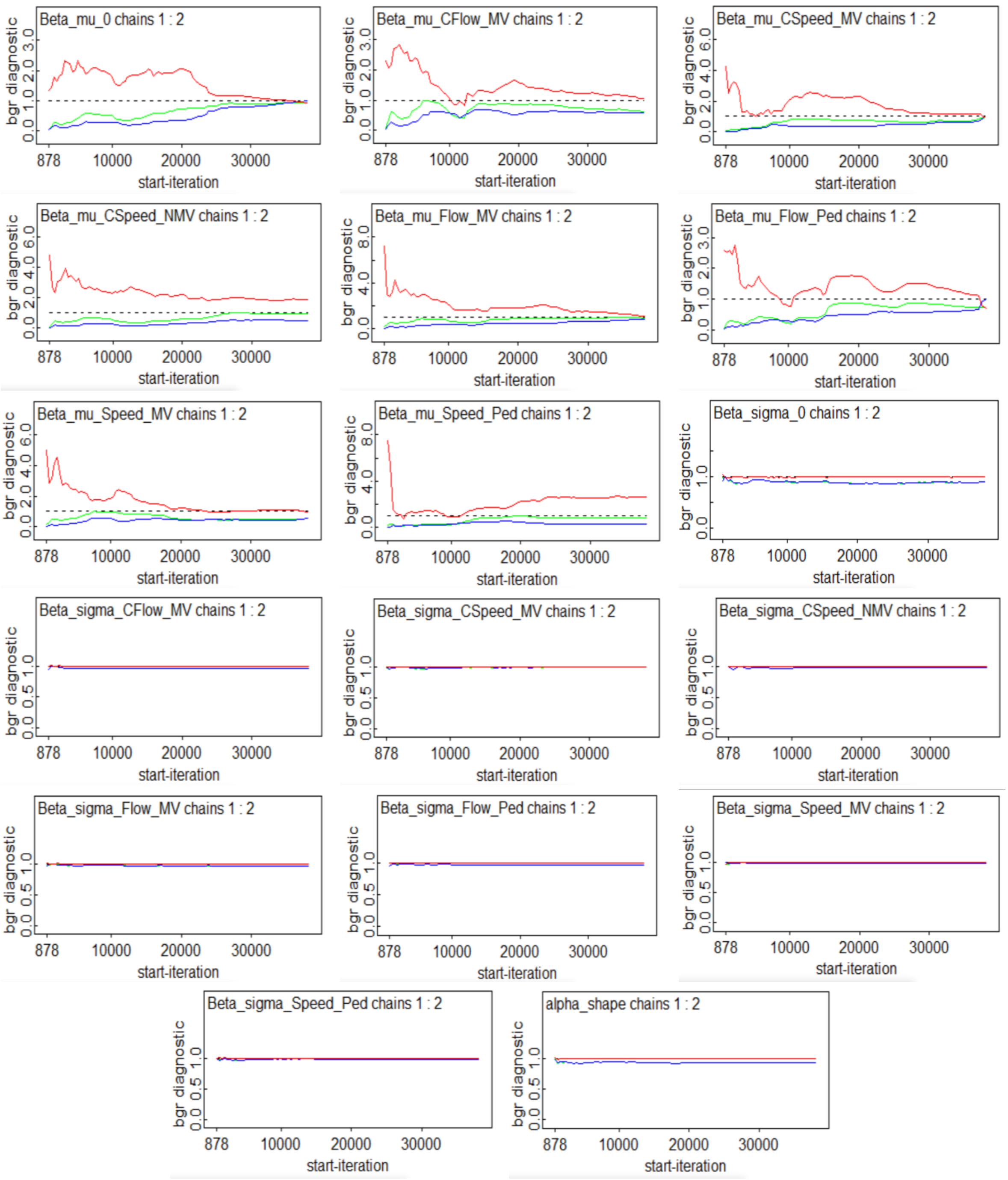

Figure A5. BGR diagrams of covariates of Model 3(a)